\documentclass{article}
\usepackage{arxiv}
\usepackage{dcolumn}
\usepackage{mathptmx} 
\usepackage{multirow}
\usepackage{} 

\usepackage[utf8]{inputenc} 
\usepackage[T1]{fontenc}    
\usepackage{hyperref}       
\usepackage{url}            
\usepackage{booktabs}       
\usepackage{amsfonts}       
\usepackage{nicefrac}       
\usepackage{microtype}      
\usepackage{lipsum}
\usepackage{amssymb,amsmath,epsfig}
\newcommand{\beq}{\begin{equation}}
\newcommand{\eeq}{\end{equation}}
\newcommand{\bea}{\begin{eqnarray}}
\newcommand{\eea}{\end{eqnarray}}

\begin{document}
\title{Circular motion around a regular rotating Hayward black hole}
\author{{Saeed Ullah Khan$^1$$^2$\thanks{saeedkhan.u@gmail.com}}\,, {Jingli Ren$^{1}$$^2$\thanks{renjl@zzu.edu.cn}}\,, and {Javlon Rayimbaev$^{3}$$^{4}$$^{5}$$^{6}$\thanks{javlon@astrin.uz}} 
\vspace{0.2cm} \\\vspace{0.1cm}
$^1$Sate Key Laboratory of Power Grid Environmental Protection/Henan Academy of Big Data, Zhengzhou University, Zhengzhou 450001, China.\\
$^2$ School of Mathematics and Statistics, Zhengzhou University, Zhengzhou, Henan 450003, China.\\
$^{3}$Ulugh Beg Astronomical Institute, Astronomy Str. 33, Tashkent 100052, Uzbekistan \\ $^{4}$Akfa University, Milliy Bogh Str. 264, Tashkent 111221, Uzbekistan \\ $^{5}$Tashkent State Technical University, Tashkent 100095, Uzbekistan \\ $^{6}$National University of Uzbekistan, Tashkent 100174, Uzbekistan}
\date{}
\maketitle
\begin{abstract}
In this article, we explore the geodesics motion of neutral test particles and the process of energy extraction from a regular rotating Hayward black hole. We analyse the effect of spin, as well as deviation parameter $g$, on ergoregion, event horizon and static limit of the said black hole. By making use of geodesic equations on the equatorial plane, we determine the innermost stable circular and photon orbits. Moreover, we investigate the effective potentials and effective force to have information on motion and the stability of circular orbits. On studying the negative energy states, we figure out the energy limits of Penrose mechanism. Using Penrose mechanism, we found expression for the efficiency of energy extraction and observed that both spin and deviation parameters, contributes to the efficiency of energy extraction. Finally, the obtained results are compared with that acquired from Kerr and braneworld Kerr black holes.\\
\end{abstract}
\section{Introduction}\label{intro}
Black holes (BHs), are among the most interesting and mysterious objects of our universe. They are accountable for many engaging astrophysical phenomena such as predicting different types of observations in the vicinity of a BH: including but not limited to jets and gamma-ray bursts \cite{Berti}; gravitational time delay; gravitational bending of light; gravitational redshift; gravitation lensing etc \cite{Pradhan1}. Recently, scientists have finally succeeded in obtaining the first-ever image of a BH, using the Event Horizon Telescope observations at the centre of the $M87$ galaxy \cite{EHT}. Among other evidence, this is the strongest ever assurance to the existence of supermassive BHs, which opens up a new window onto the investigation of BHs. They can be divided into two main categories, i.e., BHs with singularities termed as singular BHs, whereas on the other hand, BHs with no singularity is called regular (non-singular) BHs. The simplest example of a regular BH is the Reissner-Nordstr\"om (RN) BH, with the replacement of internal singularities by regular centres.

It is well known that there is no robust theory of quantum gravity, which can serve as an accurate mechanism to solve singularities of BHs. The singularity of a BH is 
supposed to be a deficiency in general relativity (GR), as it can not be explained by GR itself. Fortunately, in literature, there exist some solutions of the BH models having no singularity, called the regular BHs. Bardeen \cite{Bardeen1} was the one who for the first time introduced such type of BH. Ay\'on-Beato and Garcia (ABG) \cite{Ayon-Beato,Ayon-Beato1,Ayon-Beato2} also introduced non-singular solutions of BH models in the scenario of GR coupled to non-linear electrodynamics. Among the fascinating regular rotating BHs, Hayward \cite{Hayward} presented a static spherically symmetric BH, which behaves like de Sitter metric near the origin. Besides, such types of BHs satisfies the condition of weak energy and have everywhere finite curvature invariants. In recent years, researchers have made a considerable contribution to the study of regular BHs \cite{Toshmatov1,Halilsoy,Amir1,Amir2,MSharif,Chamseddine,Narzilloev,Narzilloev1,Rayimbaev} including the so-called Bardeen, Hayward and ABG BHs. Effects of the modifications to rotating Hayward and Bardeen black holes on particle acceleration processes have been investigated in \cite{Pourhassan}. These BHs are mostly dependent on their mass, spin and a deviation parameter. The deviation parameter is utilized to measure the potential deviation from that of the standard Kerr spacetime.

In theoretical physics, combining GR with quantum theory is a demanding issue and could only be overcome by knowing the structure of singularity encountered in GR. This makes it essential to study the geodesic structure of a BH. They could convey attractive information as well as exhibit a rich structure of the background geometry. Among different types of geodesic motions, the circular one is much more attractive. Photons can be used to obtain information on motions of BHs. The study of geodesics was for the first time introduced by Chandrasekhar \cite{Chandrasekhar} for the Schwarzschild, RN and Kerr BHs. Afterwards, many researchers take part in studying the geodesic structure around BHs \cite{Mashhoon,Shibata,Stuclik3,M.Shahzadi,Khan1,Larranaga,Lightman}. Khan and Ren \cite{KhanCJP, KhanAIP} by examining the motion of dyonic charged BH deduced that both charge and dyonic charge decreases the stability of circular orbits. Leiva et al. \cite{Leiva} explored the geodesic of Schwarzschild BH in rainbow gravity. Pugliese et al. \cite{Pug1, Pug2} by examining the RN BH extrapolate the existence of circular orbit with vanishing angular momentum. Fernando \cite{Fernando} investigate geodesics motion in the background of Schwarzschild BH surrounded by quintessence. Abbas and Sabiullah \cite{Abbas} by studying the geodesics of regular Hayward BH observed that the moving massive particles along timelike geodesics are dragged towards BH. By considering timelike geodesics, the influence of magnetic field on particle dynamics is examined around Kerr-MOG (modified gravity of theory) BH \cite{Sharif}.

In 1969, Penrose \cite{R.Penrose} for the first time inaugurate an accurate technique for the extraction of energy from spinning BHs and associated with the presence of negative energy in the vicinity of ergosphere. Lasota et al. \cite{Lasota} deduced that BHs rotational energy can be extracted from any type of matter satisfying the condition of weak energy, if and only if, it can absorb its angular momentum and negative energy. Pradhan \cite{Pradhan} explored the Penrose process in the vicinity of KNTab-NUT BH and obtained a direct connection between the gain in energy and the corresponding NUT parameter. By investigating the collisional Penrose mechanism around Kerr BH with the use of rotating particles, it is concluded that energy extraction can be greater in comparison with the non-rotating case \cite{Mukherjee}. Khan et al. \cite{saeed} analysed the Penrose mechanism around braneworld Kerr BH and observed that both of the spin and brane parameters contribute to the Penrose process. By studying rotating BH in quintessential dark energy, gathered that energy extraction could be greater in case of dark energy, in comparison with spin and charge of the BH \cite{Iftikhar}.

Efficiency of the energy extraction mechanism from revolving BHs with the help of the Penrose mechanism could usually be described as the ratio between gain and input energy. Bhat et al. \cite{Bhat} found that the efficiency of the Penrose mechanism increases due to the effect of charge around KN BH, as compared to the maximum efficiency of Kerr BH. Parthasarathy et al. \cite{Parthasarathy} by examining the spinning BH under the influence of the magnetic field, conclude that the efficiency of Penrose mechanism could reach up to $100\%$ if an incoming particle breaks in the vicinity of static limit. It is figured out that in non-Kerr BH, the deformation parameter increases the efficiency of energy extraction \cite{Liu}. Toshmatov et al. \cite{Toshmatov} observed by studying Penrose process in the vicinity of a rotating regular BH, that the efficiency of process decreases as values of the electric charge increases. Investigating the Penrose mechanism under the influence of the magnetic field, it was found that, magnetic charge contributes to the efficiency of the process \cite{Dadhich}. 

This paper explores particle dynamics of circular geodesics and the Penrose mechanism within ergoregion of a regular rotating Hayward BH. In the following section, we will review the regular rotating Hayward BH, its ergoregion, horizons and static limit. In section 3, we will discuss particle dynamics, null geodesics, the effective potential and the effective force. In Section 4, we will focus on exploring the Penrose mechanism within ergoregion, together with the negative energy states and efficiency of the aforementioned technique. Finally, in the last section, we will conclude our work with concluding remarks.
\section{Regular rotating Hayward black hole}
\label{sec:2}
The spherically symmetric static spacetime geometry of the regular rotating Hayward BH, can be described by the metric in Boyer-Lindquist coordinates as \cite{Bambi}
\begin{equation}\label{H1}
	ds^2=g_{tt}dt^2+2g_{t\phi}{dtd\phi}+g_{rr}dr^2+g_{\theta \theta}d\theta^2+g_{\phi \phi}d\phi^2,
\end{equation}
where
\begin{eqnarray}\label{H2}\nonumber
	&&g_{tt}=-\left(1-\frac{2rm}{\rho^2} \right),\quad   g_{t\phi}=-\frac{2a r m\sin^{2}\theta}{\rho^2}, \quad g_{\theta\theta}=\rho^2 , \\
	&&g_{\phi \phi}={\sin^{2}\theta}\left(a^2+r^2+\frac{2a^{2}rm \sin^{2}\theta}{\rho^2} \right), \quad g_{rr}=\frac{\rho^2}{\Delta},
\end{eqnarray}
with
\begin{equation*}
	\Delta= r^2-2rm +a^2, \quad \rho^{2}=r^2+a^2 \cos^{2}\theta \, ,
\end{equation*}
and
\begin{equation*}
	m=M\frac{r^{3+\alpha} \rho^{-\alpha}}{r^{3+\alpha}\rho^{-\alpha}+g^3 r^{\beta}\rho^{-\beta}} \, .
\end{equation*}
Here are the parameters $a$ and $M$, respectively denotes rotation and mass of the BH, $g$ represents the deviation parameter, provides a deviation from the standard Kerr BH. Both $\alpha$, $\beta$ are positive real numbers. On considering the equatorial plane the mass function takes the form
\begin{equation*}\label{H3}
	m= M \frac{r^3}{r^3+g^3} \, ,
\end{equation*}
the aforementioned mass function is no longer dependent on parameters $\alpha$, $\beta$ and $\theta$. The metric \eqref{H1}, is the generalization of Kerr spacetime, while reduces to the Schwarzschild BH by putting both $g=a=0$ \cite{Schwarz} and to the standard Kerr BH by putting $g=0$ \cite{Kerr}. For non-rotating case, i.e., $a=0$, such types of BHs satisfies the condition of weak energy, whereas disobeying the condition of weak energy in rotating case. The aforementioned metric \eqref{H1} is non-singular everywhere, including $r=0$ and $\theta={\pi}/{2}$. Note that, for simplicity, we substitute the spin parameter ${\rm a} \to a/M$ and the deviation parameter ${\rm g} \to g/M$ in all figures and tables.
\subsection{Ergoregion, horizons and static limit}
In principle, rapid rotating BHs could possess ergoregion (a region lies between the event horizon and outer static limit of a BH). The ergoregion plays a crucial role in BH physics, as Hawking radiations can be analysed within this region. Another characteristic of the ergoregion is that the test particle may have negative energy with respect to an observer at infinity. Therefore, from physical point of view, the ergoregion can allow the process of energy extraction from rotating BHs, which includes the Penrose mechanism. Moreover, one must note that particles with positive energy can enter or leave the ergoregion, whereas particles with negative energy may not be able to leave the ergoregion. Particles will have negative energy within ergoregion, if and only if the angular momentum $L$ of the particle is also negative. Specifically, on considering the circular geodesics within ergoregion both $E$ and $L$ should be negative \cite{Chandrasekhar}.

The surface of no return is called the event horizon. Whereas, the static limit surface is also termed as the infinite red-shift surface, where the time-translation killing vector becomes null. Like standard Kerr BH, the regular rotating Hayward BH has two horizons, i.e., the event and Cauchy horizons. The horizons of a BH can be obtained by solving
\begin{equation}\label{eh}
	r^2-{2mr}+a^2=0.
\end{equation} 
\begin{figure*}
	\includegraphics[width=.482\textwidth]{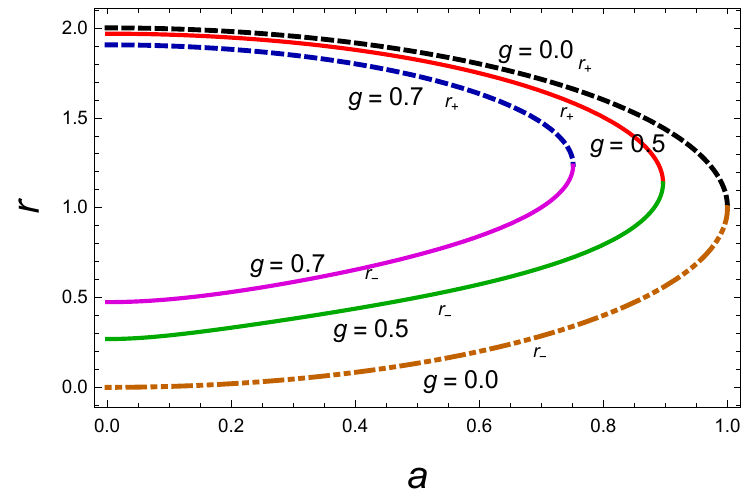}
	\includegraphics[width=0.482\textwidth]{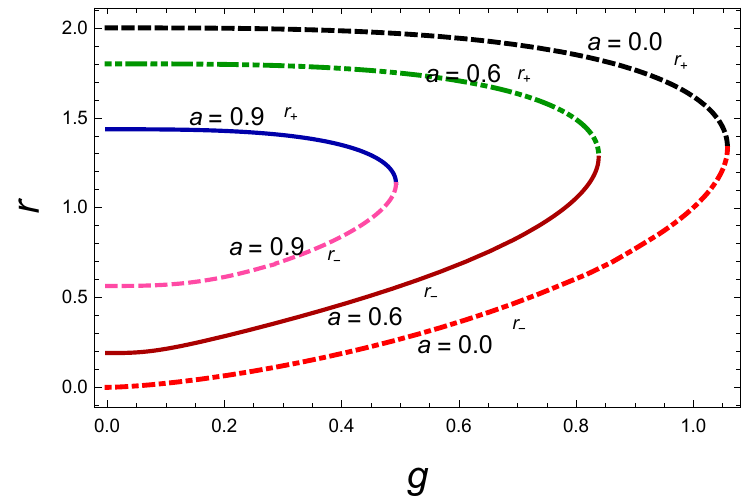}
	\caption{Graphical behaviour of the event horizon versus $a$ (left panel), whereas $g$ (right panel).}\label{Fig1}
\end{figure*}
Depending on the values of $a$ and $g$, the above equation possesses two real roots, respectively corresponds to, the event and Cauchy horizons. Among the two, the greatest possible root defines the position of the event horizon. In case, both of the horizons coincide, we obtain an extremal BH. Hence, for each non-zero value of $g$, the metric \eqref{H1} has an extremal BH. Figure \ref{Fig1}, reflects the event horizon's structure of a regular rotating Hayward, Kerr and Schwarzschild BHs. From the left panel of the graph, it is observed that in case of regular rotating Hayward BH, the event horizon becomes smaller in comparison with that of Kerr BH. In the right panel of the graph, the event horizon of Schwarzschild is larger than compared to the regular rotating Hayward BH. In addition, the event horizon gets reduce with both of the spin and deviation parameter $g$ of the BH. It is seen from this figure that an increase of the rotation and deviation parameters causes decreasing (increasing) the outer (Cauchy) horizons. It is also seen from Fig.1 in Ref.\cite{Pourhassan} for modified the Hayward BH case.

Now, we are interested in which values of the parameter $a$ and $g$ the metric (\ref{H2}) represents BH solution. In other words, we look for the values of spin and deviation parameters that serve for the existence of horizons. To do this, we set a system of equations $\Delta=0=\Delta'$ where $'$ denotes the partial derivative with respect to the radial coordinate. Below, we demonstrate the solutions of the equations graphically. 

\begin{figure*}[ht!]
	\centering
	\includegraphics[width=0.7\textwidth]{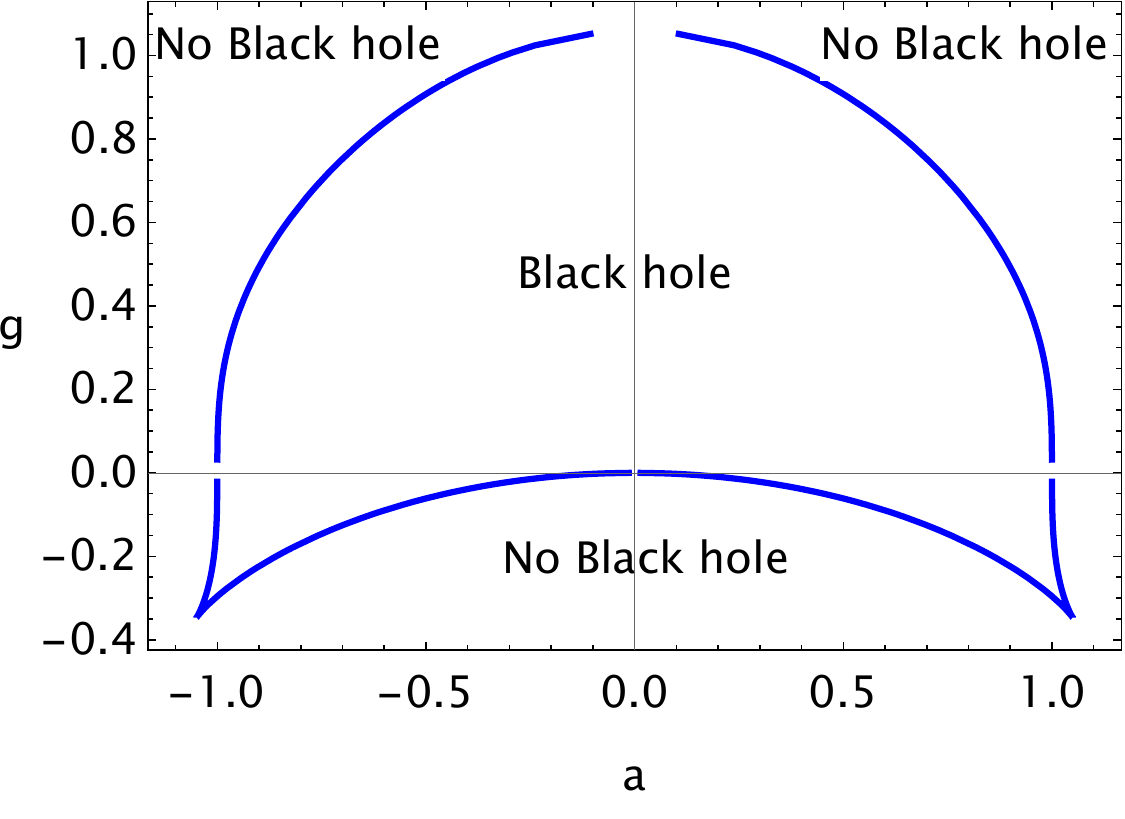}
	\caption{Relationship between the spin $a$, and deviation parameter $g$, for the existence of BH horizon.} \label{avsg}
\end{figure*}

In Fig.~\ref{avsg} we present values of parameters of spin of Hayward BHs and deviation. 
The area inside the blue contour indicates that the values of these parameters correspond to the existence of BH horizons, while the blue line represents their extreme values. This means that there is no horizon(s) for the values of spin and deviation parameters outside the blue contour.

The static limit surface also has interesting and important consequences in BH physics. The key property of this surface is to change the nature of particle geodesics by crossing this surface. On substituting $g_{tt} = 0$, the static limit surface can be obtained using the following equation
\begin{equation}\label{sls}
	r^2-{2mr}+a^2 cos^{2}\theta=0.
\end{equation}
\begin{figure*} \vspace{-0.0cm}
	\includegraphics[width=0.95\textwidth]{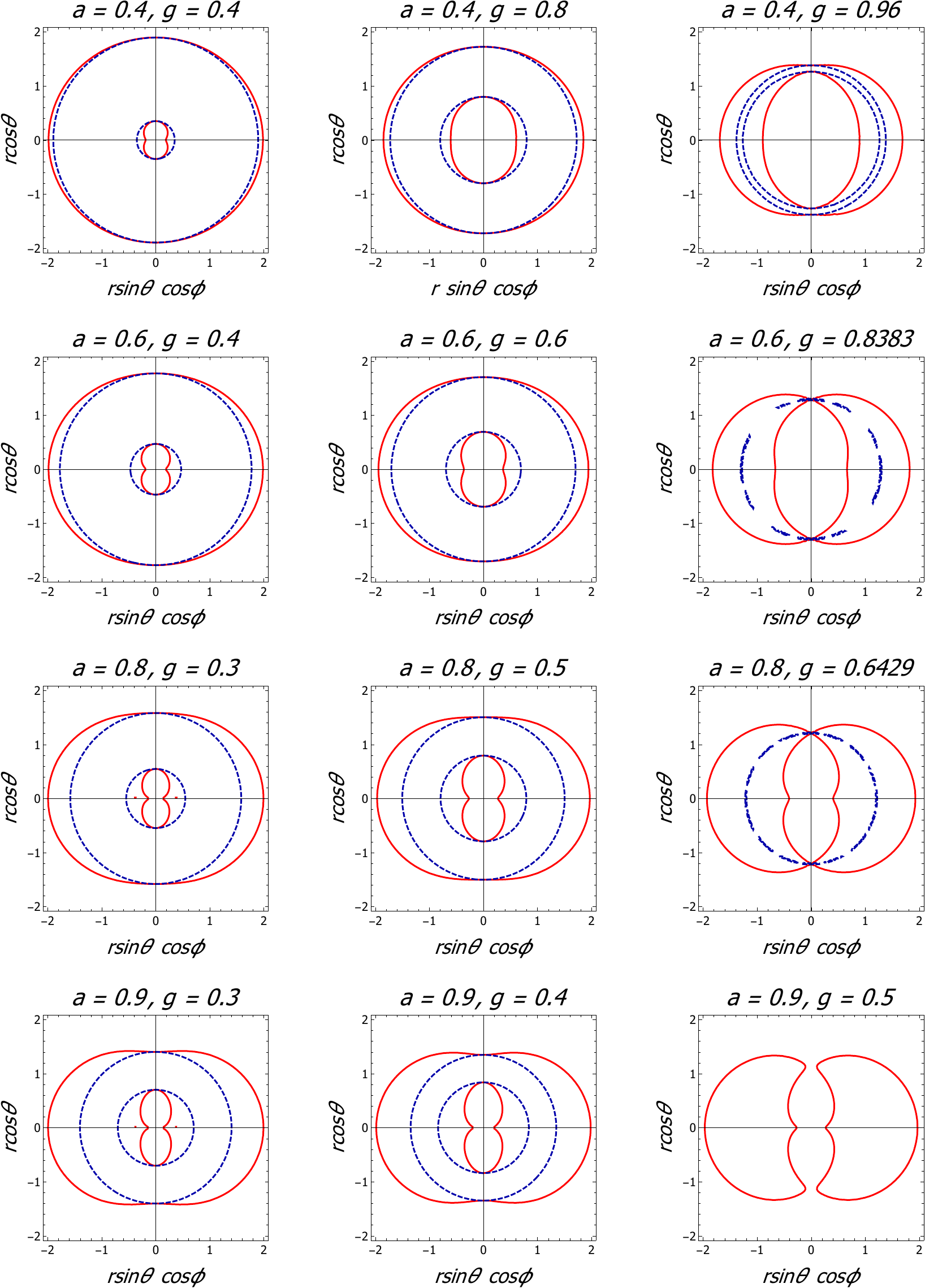}
	\caption{Graphical representation of the ergoregion, static limit (red solid curves) and event horizon (blue dotted curves) in the $x-z$ plane at various discrete points of $g$ and $a$.} \label{ergo1}
\end{figure*}
\begin{table}\label{Tab1}
	\begin{center}
		\textbf{Table 1:} The event horizon ($r_+$), \\static limit ($r_{+es}$) and ergoregion ($r_{+es}-r_+=\delta$) of the BH.
		\\
		\small\addtolength{\tabcolsep}{0.pt}
		\scalebox{0.98}{
			\begin{tabular}{|c| c| c| c| c| c|}
				\hline  \noalign {\smallskip}
				&  {a=0.5}  &  {a=0.7}  &   {a=0.8} & {0.9} & {a=0.95}  \\[.5em]
				\hline \noalign {\smallskip}
				$g$ & $r_+ \quad \quad r_{+es} \quad \quad \delta$ & $r_+ \quad \quad r_{+es} \quad \quad \delta$ & $r_+ \quad \quad r_{+es} \quad \quad \delta$ $b$ & $r_+ \quad \quad r_{+es} \quad \quad \delta$ & $r_+ \quad \quad r_{+es} \quad \quad \delta$   \\[.5em]
				\hline \noalign {\smallskip}
				
				0.0 &  1.866  1.935 0.069& 1.714  1.869  0.155& 1.600 1.825  0.225& 1.436 1.771  0.335& 1.312  1.741  0.429\\[.5em]
				
				0.2 &  1.863  1.933  0.070& 1.710  1.866 0.156& 1.595 1.822  0.227& 1.427 1.768  0.341& 1.297  1.737 0.440  \\[.5em]
				
				0.4 &  1.844  1.917  0.073& 1.682  1.847 0.165 &1.555 1.801  0.246& 1.348 1.744  0.396& \quad - \quad 1.711  \quad - \quad \\[.5em]
				
				0.6 &  1.787  1.869  0.082& 1.590  1.791 0.201& 1.392 1.737  0.345&  \quad - \quad 1.668  \quad - \quad & \quad - \quad  1.627 \quad  - \quad \\[.5em]
				\hline 
		\end{tabular}}
	\end{center}
\end{table}
It can be observed from Eqs. \eqref{eh} and \eqref{sls}, that the parameters $a$ and $g$ have direct effects on horizons, as well as on the static limit surface of the BH. Moreover, for each value of the deviation parameter $g$, there exists a critical value of $a$ and $r$ \cite{Amir2}. The behaviour of ergoregion, event horizon and static limit surface, is revealed in Fig. \ref{ergo1} and in Table {\bf1}, at various discrete points of the spin and deviation parameter $g$, of the BH. It is found that both spin and deviation parameter $g$, has a great influence on the geometrical shape of ergoregion, event horizon and static limit. Moreover, it is observed that both $a$ and $g$, results in decreasing the radii of static limit, as well as of the event horizon. Furthermore, both $a$ and $g$ results in increasing the ergoregion's area, while an observer within the ergoregion can not remain stable. Henceforth, each value of the rotation parameter $a$ ensures a critical value $g=g_c$ provides that for $g<g_c$, results in non-extremal Hayward BH with only two horizons. We obtain an extremal BH at $g=g_c$, with two coincide horizons, whereas in case of last possibility $g>g_c$ no horizons or BH, could be obtained.
\section{Particles motion and the innermost stable circular orbit}
\label{sec:3}
The current section is devoted to the study of particle dynamics, which is very essential to explore negative energy states and the efficiency of Penrose mechanism. To obtain geodesics in the equatorial plane ($\theta=\pi/2$), we will utilize neutral particle motion around a regular rotating Hayward BH. Therefore, using the Lagrangian equation, the equation governing geodesics could be obtained as
\begin{equation}
	\mathcal{L} = \frac{1}{2}g_{\mu\nu}\dot{x}^{\mu}\dot{x}^{\nu},
\end{equation}
here $\dot{x}^{\mu}$ denotes the four-velocity and dot represent $\partial / \partial \tau$ (with $\tau$ is the proper time). The generalized momenta for \eqref{H1}, could be expressed as
\begin{eqnarray}\label{6}
	p_{t}&=&g_{tt}\dot{t}+g_{t\phi}\dot{\phi}=-E,\\\label{7}
	p_{\phi}&=&g_{t\phi}\dot{t}+g_{\phi\phi}\dot{\phi}=L,
	\\\nonumber p_{r}&=&g_{rr}\dot{r}.
\end{eqnarray}
In Eqs. \eqref{6} and \eqref{7} $E$ and $L$, represents the total energy and angular momentum of the test particle, respectively related to the Killing vector fields $\xi_{t}=\partial_{t}$ and $\xi_{\phi}=\partial_{\phi}$. Since along geodesics both of the $p_{t}$ and $p_{\phi}$ are preserved and thus describes spherically symmetric and static properties of the regular rotating Hayward BH, as the Lagrangian does not depend on $t$ and $\phi$. Solving Eqs. (\ref{6}) and (\ref{7}), for $\dot{t}$ and $\dot{\phi}$,  we get
\begin{eqnarray}\label{energy}
	\dot{t}&=&\frac{1}{r^2 \Delta}\left(a^2 E(2mr + r^2) +E r^4-2 a mLr\right),\\\label{angmomentum}
	\dot{\phi}&=&\frac{1}{r^2 \Delta}\left(L(r^2-2mr) +2aEmr\right)\ .
\end{eqnarray}

For the neutral particle motion, the Hamiltonian takes the form
\begin{equation}\nonumber
	H=p_{t}\dot{t}+p_{r}\dot{r}+p_{\phi}\dot{\phi}-\mathcal{L}.
\end{equation}
In case of metric \eqref{H1}, the aforementioned equation could be rewritten as
\begin{eqnarray}\nonumber
	2H&=&\left(g_{tt}\dot{t}+g_{t\phi}\dot{\phi}\right)\dot{t}+
	\left(g_{t\phi}\dot{t}+g_{\phi\phi}\dot{\phi}\right)\dot{\phi}+g_{rr}\dot{r}^{2}\\\label{hamiltonian}
	&=&-E\dot{t}+L\dot{\phi}+\frac{r^{2}}{\Delta}\dot{r}^{2}=\epsilon=\text{constant},
\end{eqnarray}
here $\epsilon=-1, 0, 1$, respectively, denote the timelike,  lightlike (null) and
spacelike geodesics. By inserting the values of Eqs. \eqref{energy} and \eqref{angmomentum} into \eqref{hamiltonian}, the radial equation of motion can be obtained as
\begin{eqnarray}
	\label{radial}
	\dot{r}^{2}r^{4}=E^{2}r^{4}+(a^{2}E^{2}-L^{2})r^{2}+ \Delta r^{2} \epsilon +2(aE-L)^{2}m r.
\end{eqnarray} 

The above obtained results, in Eqs. \eqref{energy}-\eqref{radial}, are very essential, as they could be used to investigate the associated properties of particle dynamics around regular rotating Hayward BH.

To investigate the geodesics motion, we describe the right-hand side of Eq. \eqref{radial}, by $\mathcal{R}(r)$
\begin{equation}\label{ISCO1}
	\mathcal{R}(r)=E^{2}r^{4}+(a^{2}E^{2}-L^{2})r^{2}+ \Delta r^{2} \epsilon +2(aE-L)^{2}mr.
\end{equation}
In case of circular orbits, the following two constraints must be holds simultaneously
\begin{equation}\label{ISCO2}
	\mathcal{R}(r)=0, \quad \partial_r \mathcal{R}(r)=0.
\end{equation}
Henceforth, using the above constraints and spacetime line element of the regular Hayward BH. The specific energy ${E}$ and angular momentum $L$, related to a distant observer at a given radius $r$, can be obtained as \cite{Aliev, Stuch4}
\begin{eqnarray}\label{ISCO3}
	&&\frac{E}{\epsilon}=\frac{\Delta-a^2\pm a\sqrt{mr}}{r\sqrt{r^2-3mr\pm 2a\sqrt{mr}}},\\\label{ISCO4} 
	&&\frac{L}{\epsilon}=\pm\frac{\sqrt{mr} \left(a^2+r^2\mp 2a\sqrt{mr}\right)}{r\sqrt{r^2-3mr\pm 2a\sqrt{mr}}}.
\end{eqnarray}
Here the upper and lower signs, at a larger radial distance $r$, respectively correspond to the co-rotating and counter-rotating orbits.
It can be observed from Eqs. \eqref{ISCO3} and \eqref{ISCO4} that, the existence of circular geodesics needs
\begin{eqnarray}\label{ISCO5}
	{r^2-3mr\pm 2a\sqrt{mr}} \geq 0.
\end{eqnarray}
While, the stability of circular orbits requires that
\begin{equation}\label{ISCO6}
	\partial_{rr}{\mathcal{R}(r)} \geq 0.
\end{equation}
In Eq. \eqref{ISCO6}, equality determines the innermost stable circular orbits (ISCO). Since the stability of circular orbits requires $r_{ISCO}<r$, implies that $r$ can not be too small.  Therefore, using the condition $\partial_{rr}\mathcal{R}(r)=0$, the relation of ISCO can be obtained as
\begin{eqnarray}\label{ISCO7}
	\left(3a^2+6mr_{ISCO}-r_{ISCO}^2\right)mr_{ISCO} - 8a (mr_{ISCO})^{3/2}=0.
\end{eqnarray}
On substituting $a=g=0$, the metric \eqref{H1} reduces to the Schwarzschild BH and $r_{ISCO}=6M$, while for $g=0$ and $a=M$, we obtain $r_{ISCO}=M$.

\begin{table}[b]\label{Tab22}
	\begin{center}
		\textbf{Table 2:} Radius of the event horizon ($r_+$), and ISCO ($r_{ISCO}$) around the BH.
		\\
		\small\addtolength{\tabcolsep}{0.pt}
		\scalebox{0.78}{
			\begin{tabular}{|c| c| c| c| c| c|c|}
				\hline  \noalign {\smallskip}
				a $\to$&  {0.3} &  {0.5}  &  {0.7}  &   {0.8} & {0.9} & {0.95}  \\[.5em]
				\hline \noalign {\smallskip}
				$g \ \downarrow$ & $r_+ \quad \quad r_{ISCO}$ & $r_+ \quad \quad r_{ISCO}$ & $r_+ \quad \quad r_{ISCO}$ & $r_+ \quad \quad r_{ISCO}$ & $r_+ \quad \quad r_{ISCO}$ & $r_+ \quad \quad r_{ISCO}$  \\[.5em]
				\hline \noalign {\smallskip}
				
				0.0 &  1.9539 \quad  4.9786 &  1.866 \quad  4.2330 & 1.714 \quad  3.3931  & 1.600 \quad 2.9066 & 1.436 \quad  2.3209& 1.312\quad 1.9372\\[.5em]
				
				0.2&  1.9517 \quad  4.9782 &  1.863 \quad   4.2323& 1.710 \quad  3.3917& 1.595 \quad  2.9043 & 1.427 \quad  2.3155& 1.297 \quad  1.9255  \\[.5em]
				
				0.4 &  1.9361 \quad  4.9755&  1.844 \quad  4.2277& 1.682 \quad  3.3822 &1.555 \quad   2.8878 & 1.348 \quad   2.2752& \quad - \quad   \quad - \quad \\[.5em]
				
				0.6 &  1.8904 \quad  4.9679&  1.787 \quad  4.2152& 1.590 \quad  3.5560& 1.392 \quad  2.8402&  \quad - \quad  \quad - \quad & \quad - \quad  \quad  - \quad \\[.5em]
				\hline 
		\end{tabular}}
	\end{center}
\end{table}

Radius of outer event horizon of rotating Hayward black holes and ISCO around the BH for test particles are presented in Table~2. It is observed from this table that the ISCO radius decreases rapidly with the increase of the spin of the Hayward BHs and comes close to the BH's outer horizon, while it slight decreases with respect to the increase of the deviation parameter.

\begin{figure*}
	\includegraphics[width=.495\textwidth]{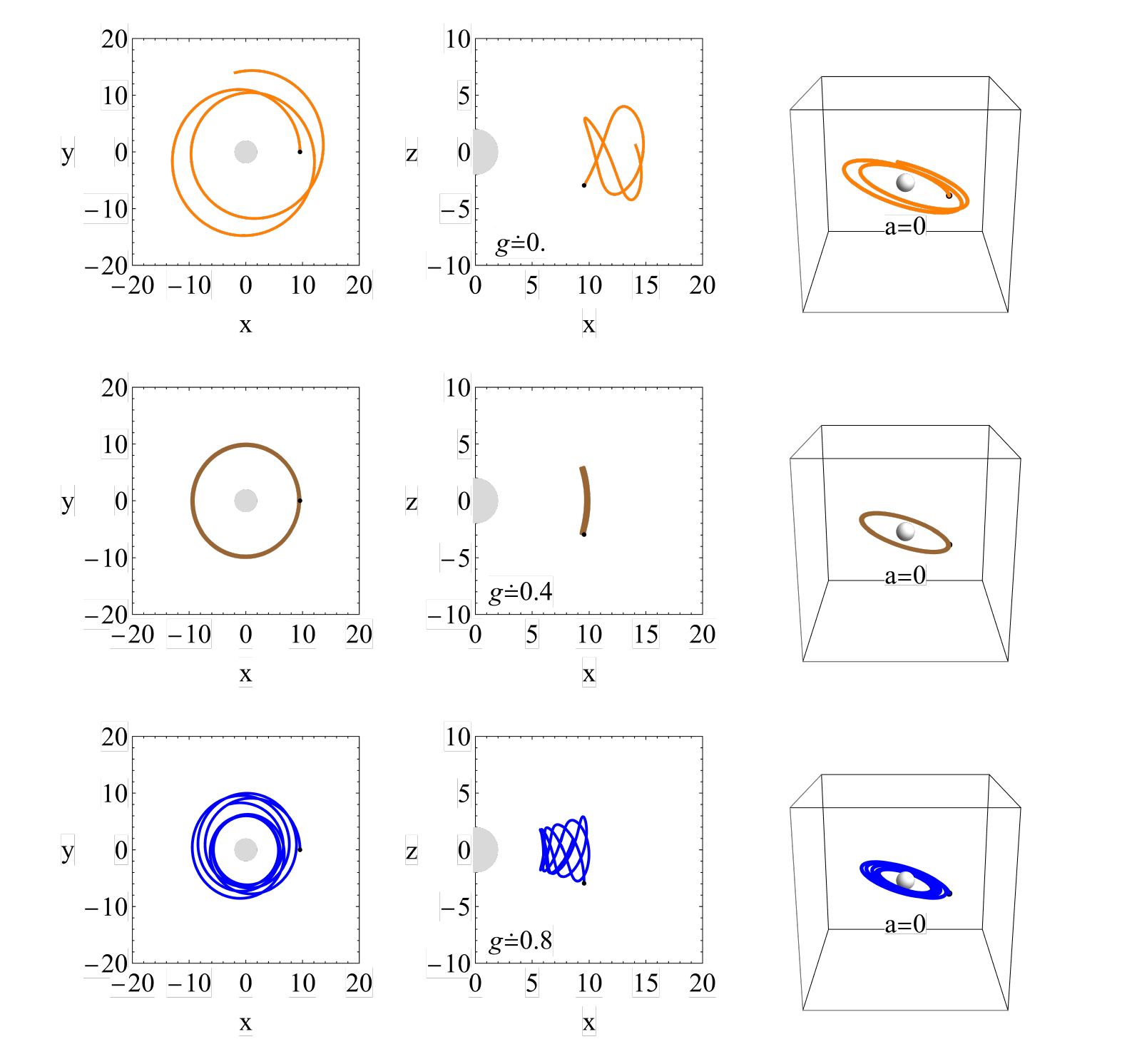}
	\includegraphics[width=0.495\textwidth]{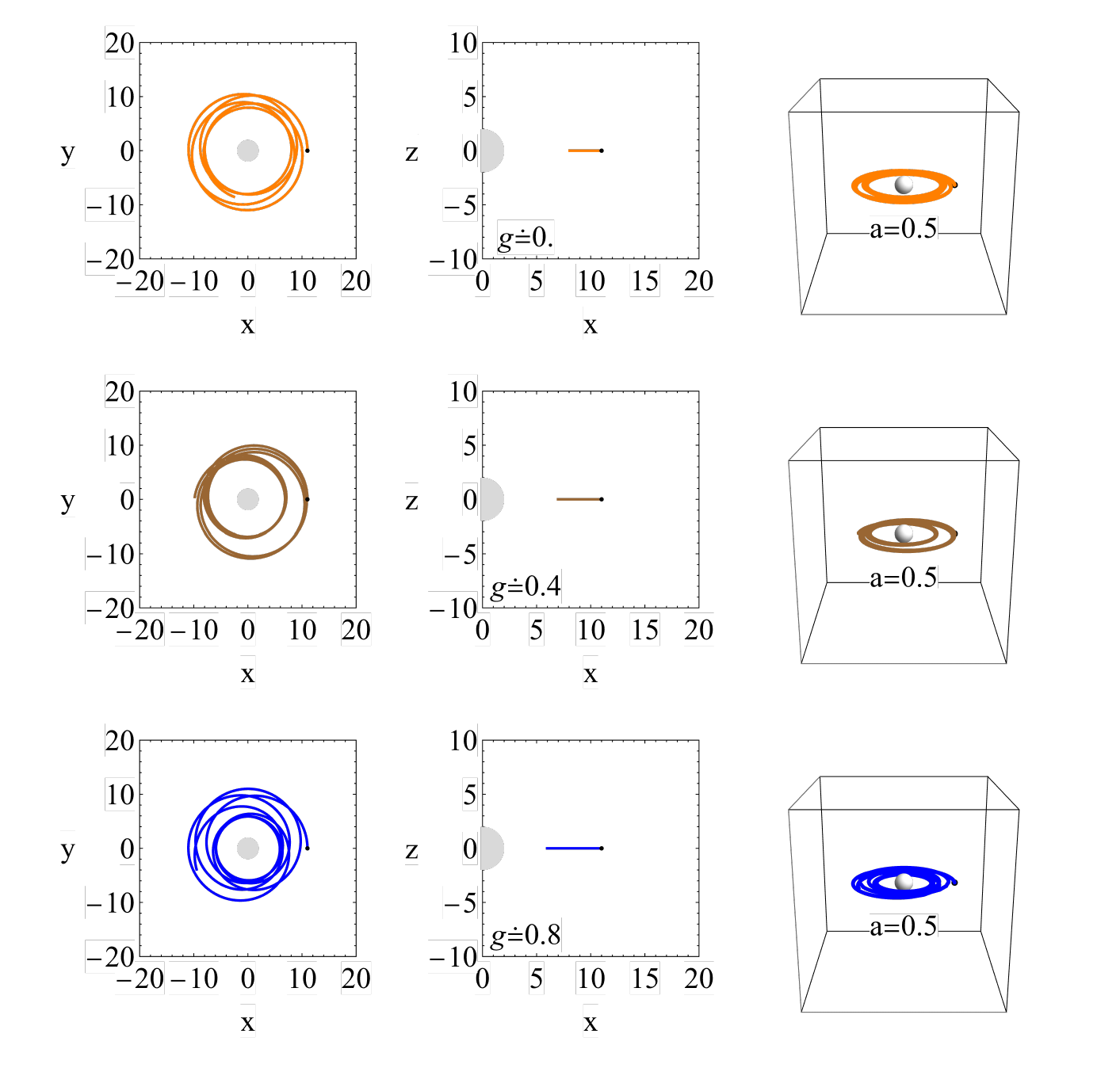}
	\caption{Trajectories of test particles around stating (left panel) and rotating (right panel, $a/M=0.5$) Hayward BHs with different values of the deviation parameters ($g/M$=0.4 in middle row and $g/M$=0.8 in bottom row). In the left panel we select the initial coordinates $r_0=10M$ \& $\theta_0=\pi/2+0.2$, while in the right panel $r_0=11M$ \& $\theta_0=\pi/2$. In both cases angular momentum is choisen as $L^2=14$.}\label{Tr}
\end{figure*}

It is observed from Fig.\ref{Tr} that the trajectory particles come more close as the deviation parameter is increased. While, with an increase of BH spin, forces the particles to rotate around the BH rapidly at the same proper time $\tau$ making denser paths.

\subsection{Null geodesics}
The radial Eq. \eqref{radial}, for lightlike particles moving along geodesics, can be rewritten as

\begin{eqnarray}\label{ng1}
	\dot{r}^2= E^{2}+\frac{1}{r^{2}}(a^{2}E^{2}-L^{2})+\frac{2}{r^{4}}(aE-L)^{2}m r.
\end{eqnarray}

For differentiating the geodesics, it would be helpful to inaugurate an impact parameter $D=L/E$, in place of $L$. Thus, we assume a special case, i.e., $L=aE$, where  $D=a$. Henceforth, for $\dot{t}$, $ \dot{\phi}$ and $\dot{r}$, respectively, we get the following expressions
\begin{eqnarray}\label{ng2}
	\dot{t}=\frac{(a^2+r^2)E}{\Delta} ,\quad \dot{\phi}=\frac{aE}{\Delta} \quad  \text{and} \quad \dot{r}=\pm{E}.
\end{eqnarray}
Here the dot indicates the derivative with respect to the affine parameter $\tau$, while the $\pm$ signs, respectively, represents the outgoing and incoming photons. The corresponding expression for $t$ and $\phi$ could be obtained as
\begin{eqnarray}\label{ng3}
	\frac{dt}{dr}=\pm \frac{(a^2+r^2)}{\Delta}, \quad \frac{d \phi}{dr}=\pm \frac{a}{\Delta}. 
\end{eqnarray}
Our interest lies in the general case when $aE \neq L$ and finds out the radius of unstable photon orbit where $E=E_c$, $L=L_c$ and $D_{c}=L_c/E_c$. Henceforth, Eq. \eqref{ng1} and its derivative simplifies to
\begin{equation}\label{ng4}
	r_{c}^{2}+(a^2-D_{c}^{2})+\frac{2(a-D_{c})^2 m r_c}{r_c^{2}}=0,
\end{equation}
\begin{equation}\label{ng5}
	r_{c} -\frac{3M r_c^{4}(a-D_{c})^2}{(g^{3}+r_{c}^3)^2}+\frac{2Mr_c(a-D_{c})^2}{(g^3+r_{c}^3)}=0.
\end{equation}
Using the aforementioned equation we can obtain
\begin{equation}\label{ng6}
	D_{c}=a \mp \sqrt{\frac{(r_{c}^3+g^3)^2}{M{(r_{c}^3-2g^3)}}}.
\end{equation}
By inserting Eq. \eqref{ng6} into \eqref{ng4}, the equation of circular photon orbit can be acquired as
\begin{align}\label{ng7}
	(r_c^3+g^3)-2Mr_c^2- Mr_c^2 {\left(\frac{r_c^3-2g^3}{r_c^3+g^3}\right)} \pm 2a \sqrt{M(r_c^3-2g^3)}=0.
\end{align}
To obtain the radius of photon orbit for regular rotating Hayward BH, we assume that $r_{c}=r_{ph}$, be the real positive root of Eq. \eqref{ng7}. This is the closest possible circular orbit of a photon at $r=r_{ph}$, to the regular Hayward BH. The photon orbits of Kerr BH can be recovered by substituting $g=0$ \cite{Bardeen}, whereas by putting $a=g=0$, we can recover photon orbits of the Schwarzschild BH. 
\begin{figure*}
	\includegraphics[width=.482\textwidth]{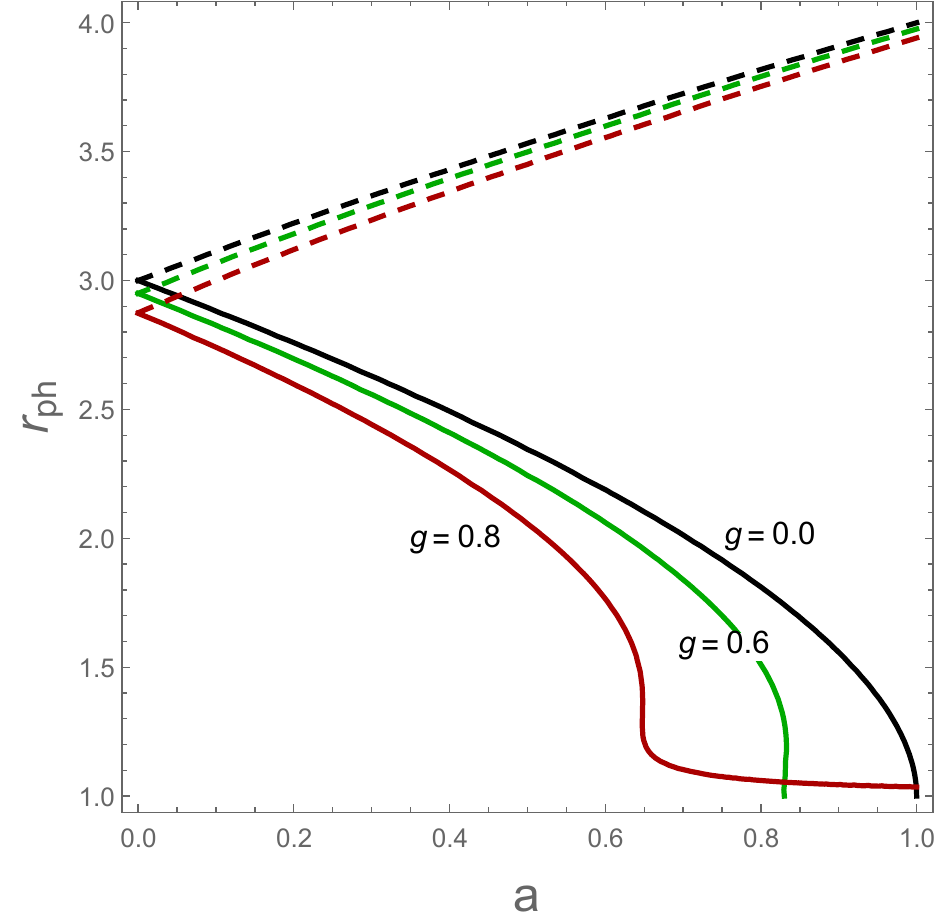}
	\includegraphics[width=0.482\textwidth]{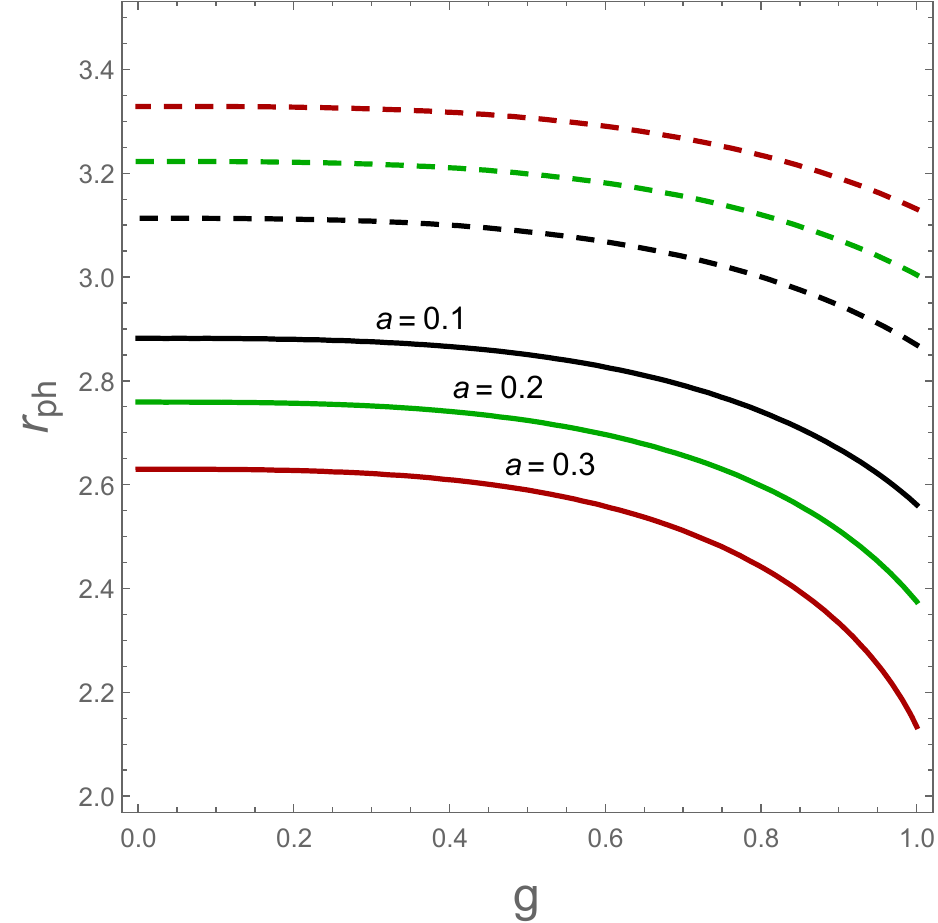}
	\caption{Graphical descriptions of photon orbits $r_{ph}$, as a function of $a$. The left channel is for prograde orbits, while the right channel is for retrograde orbits.}\label{ph}
\end{figure*}

The affected behaviour of photon orbits by the deviation parameter $g$ and spin $a$, of the BH is shown in Fig \ref{ph}. The solid curves describe co-rotating, while the dashed curves denotes counter-rotating motion of particles. We observed that the deviation parameter $g$, result in diminishing the photon orbits of both prograde and retrograde motion whereas BH spin diminishing the photon orbits of prograde motion, while contributes to the photon orbits in case of retrograde motion. By making use of Eqs. \eqref{ng4} and \eqref{ng6}, we obtain
\begin{equation}\label{ng8}
	D_{c}^2=a^2+\frac{3r_{c}^5}{r_{c}^3-2g^3} .
\end{equation}
An essential physical quantity known is the angular frequency $\Omega_{c}$ (associated with circular null geodesics) observed by an asymptotic observer can be derived as
\begin{equation}\label{ng9}
	\Omega_c=\frac{\left[(r_c^2-{2mr_c})D_c+{2am r_{c}}\right]}{\left[r_c^2(r_{c}^2+a^2)+{2a^2m r_{c}}-{2amr_c D_{c}}\right]} =\frac{1}{D_{c}}.
\end{equation}
It can be observed from Eqs. \eqref{ng4} and \eqref{ng6}, that the impact parameter is the inverse of the aforementioned angular frequency and generalizes the result of Kerr BH \cite{Chandrasekhar} to the regular rotating Hayward BH.
\subsection{Effective potential}
The effective potential ($U_{eff}$), can be utilized to determine the range of angular momentum and to discuss the stability of circular orbits of a particle. The stable and unstable orbits could be determined by the use of maximum and minimum values of $U_{eff}$, respectively. Whereas, the supreme value of $U_{eff}$ can be obtained at $r=0$. The radial Eq. \eqref{ng1}, for, $L \neq aE$ can be expressed as \cite{Misner,Abdujabbarov}
\begin{equation}\nonumber
	\dot{r}^{2}=E^2- U_{eff}.
\end{equation}
\begin{figure*} 
	\includegraphics[width=0.482\textwidth]{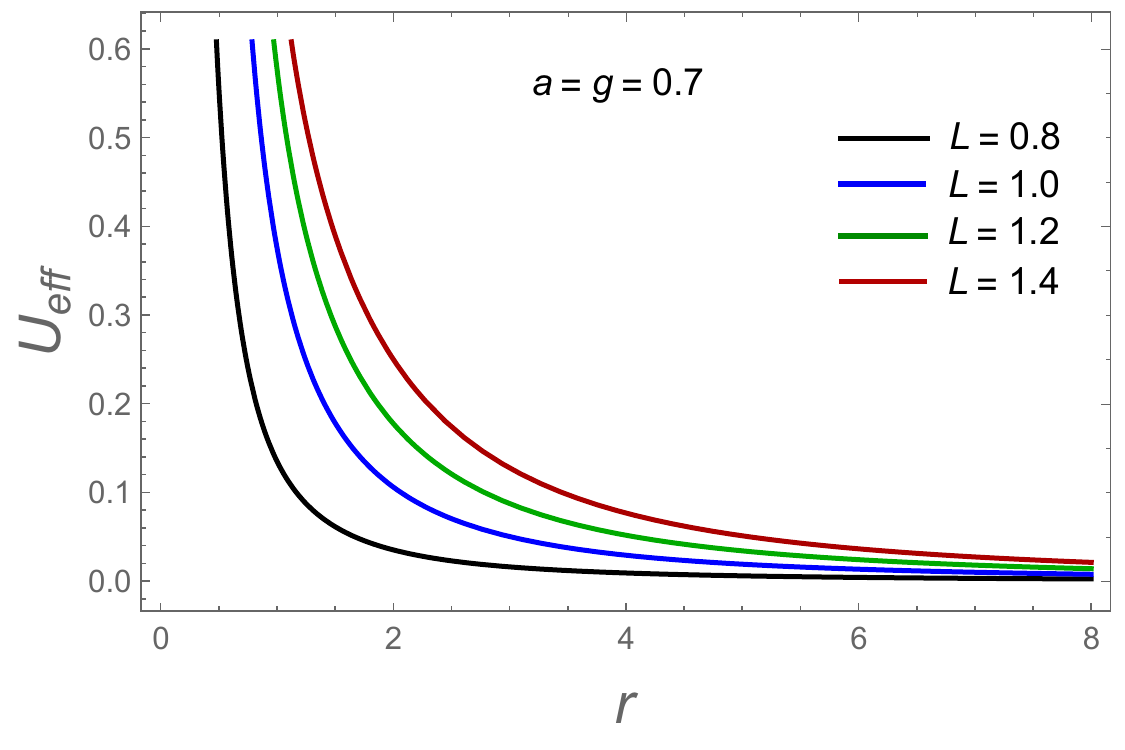}
	\includegraphics[width=0.482\textwidth]{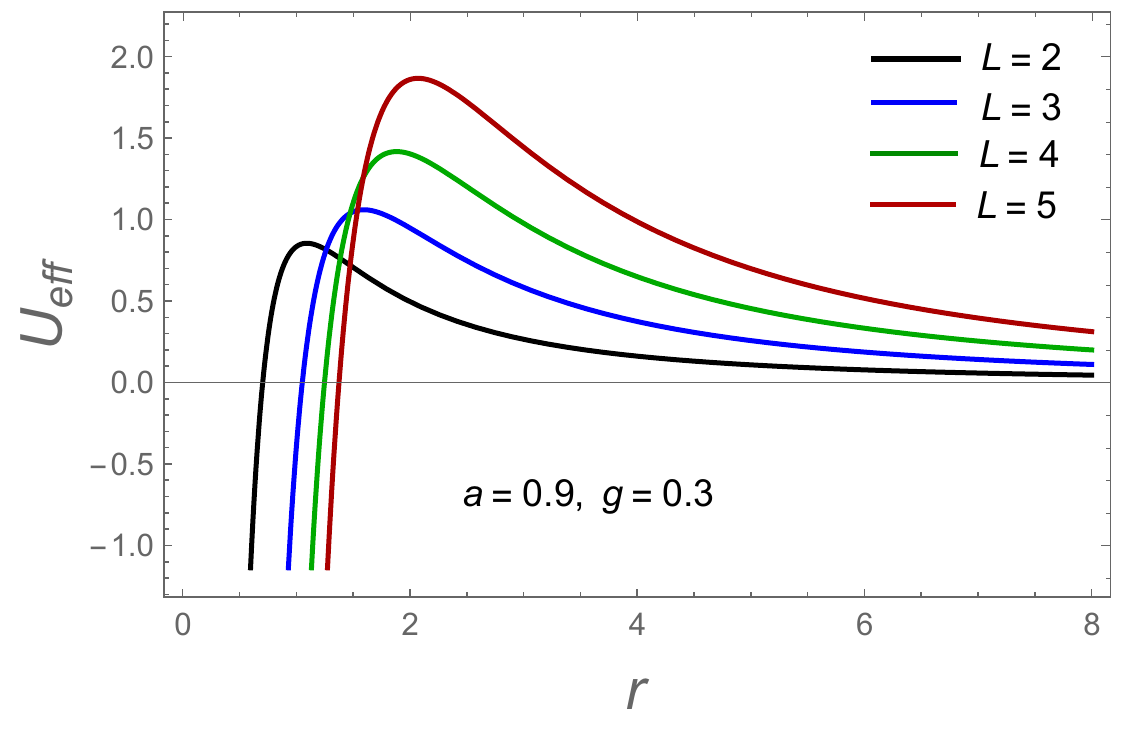}
	\includegraphics[width=0.482\textwidth]{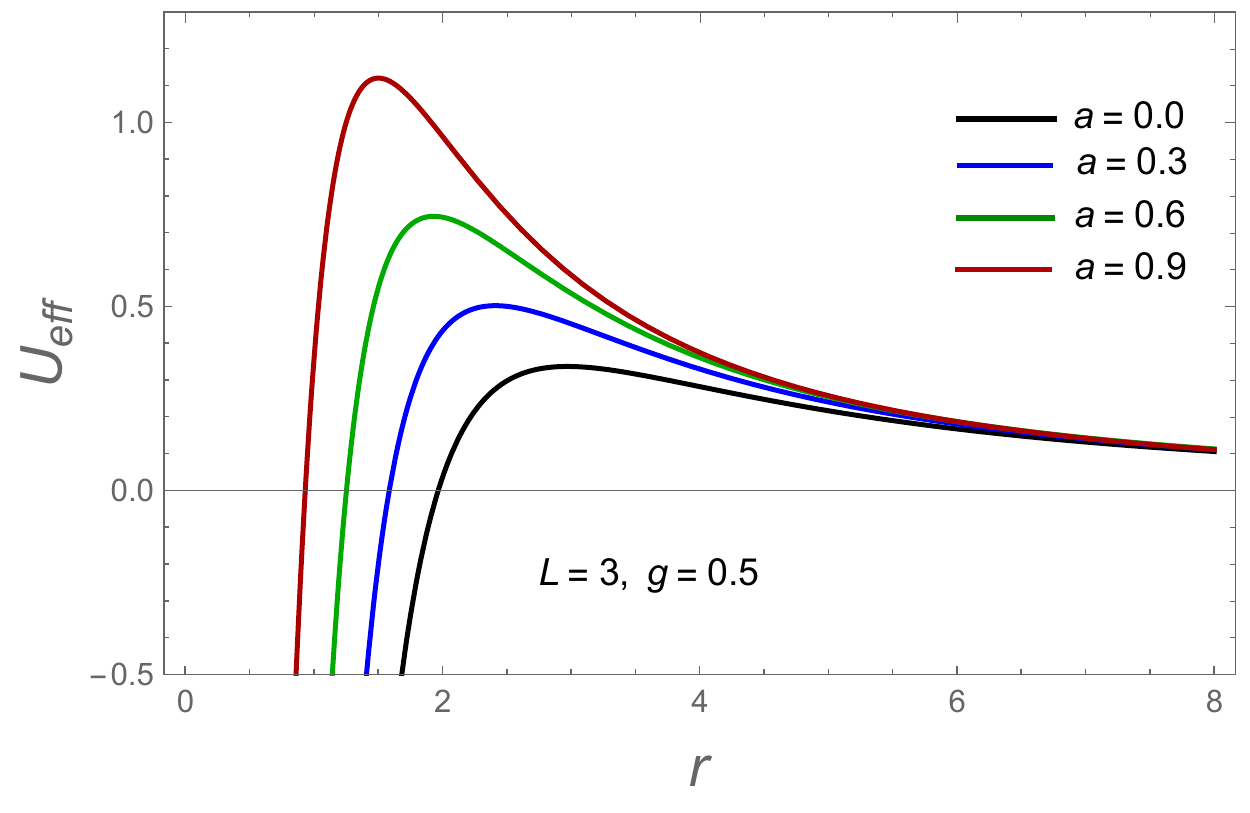}
	\includegraphics[width=0.482\textwidth]{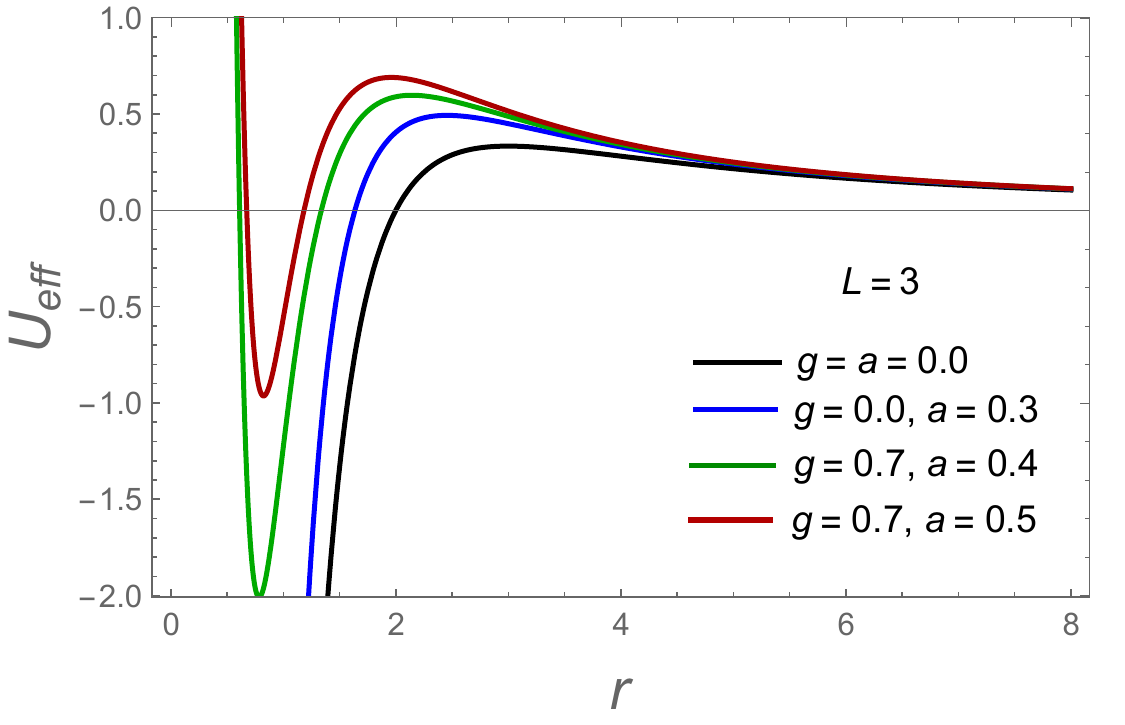}
	\caption{Plots showing the behaviour of $U_{eff}$, as a function of $r$ with the assumption of $E=1$.}\label{EP}
\end{figure*}
As a result, the effective potential takes the form
\begin{equation} \label{eff1}
	U_{eff}=\frac{(L^2-a^2E^2)}{r^2}-\frac{2(L-aE)^2m}{r^3}.
\end{equation}
The initial radial acceleration, as well as the radial velocity, should be vanished for a particle to describe the circular orbits. 
While in case of a stable circular orbit, the effective potential should be minimum
\begin{equation*}
	\frac{\partial{^2} U_{eff}}{\partial r^{2}}>0.
\end{equation*}
The graphical behaviour of $U_{eff}$ of lightlike particles moving around a regular rotating Hayward BH is depicted in Fig. \ref{EP}, at different values of the parameters $a$, $g$ and $L$.
The upper row left, and right panels are plotted at various values of $L$ and $a$, respectively. From where it can be observed that BH spin increases the instability of circular orbits. In the bottom row, the left panel $U_{eff}$ is plotted at different values of the angular momentum $L$, which shows that $L$ increases the instability of circular orbits and attains its maximum values  $U_{eff} \approx 0.85, 1.05, 1.41, 1.86$, respectively at $L = 2, 3, 4 ,5$. The bottom right panel shows that circular orbits around a regular rotating Hayward BH is initially stable but become more unstable along with radial distance $r$, as compared to both of the Schwarzschild and Kerr BHs.
\subsection{Effective force}
The study of effective force is of great interest, as it could be used to investigate the information of motion. It also let us know whether the particle is moving toward or away from the BH. Using Eq. \eqref{eff1}, the effective force on a photon could be determined as \cite{S. Fernando}
\begin{align}\label{EF1} 
	\mathcal{F}=-\frac{1}{2}\left( \frac{d U_{eff}}{dr}\right) = \frac{(L^2-a^2 E^2)}{r^3}-\frac{3 m (L-aE)^2}{r \left(g^3+r^3\right)}.
\end{align}
\begin{figure*}\vspace{-0.5cm}
	\includegraphics[width=0.482\textwidth]{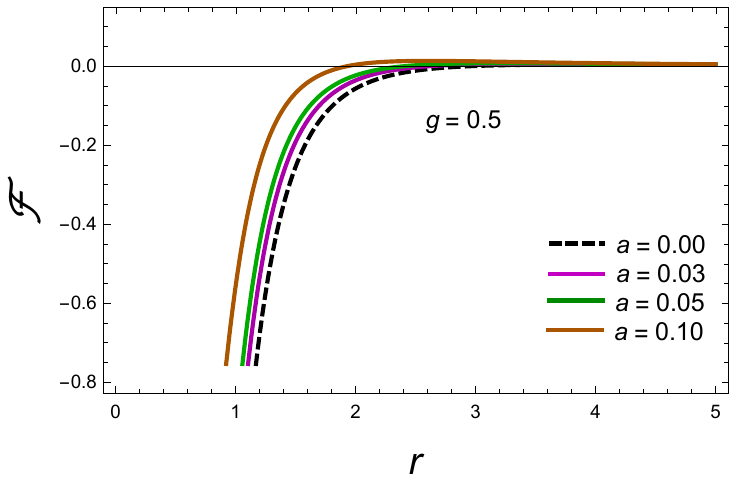}
	\includegraphics[width=0.482\textwidth]{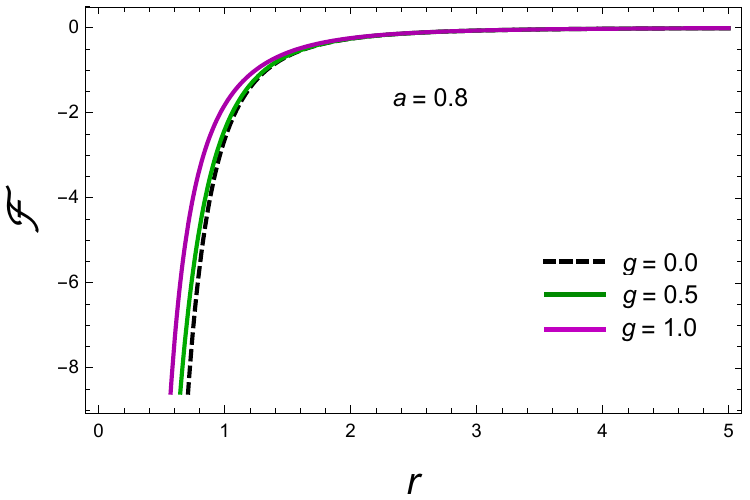}
	\centering
	\includegraphics[width=0.482\textwidth]{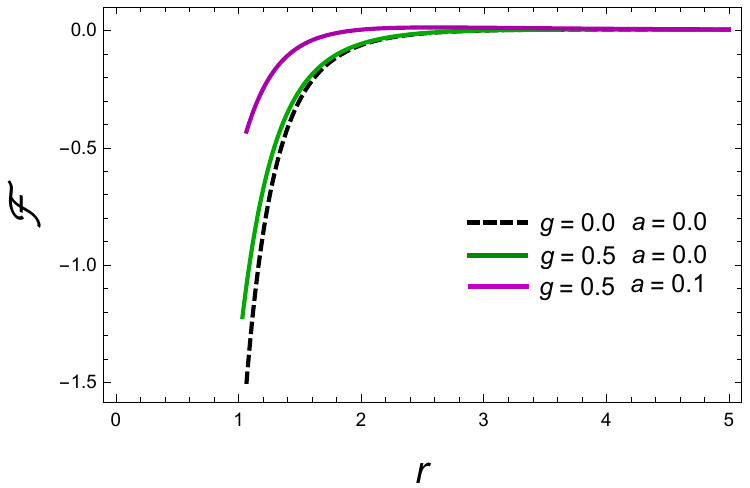}
	\caption{Graphical representation of the effective force $\mathcal{F}$, as a function of $r$, with the assumption of $E=2$ and $L=1$.}\label{EF}
\end{figure*}
It can be observed that the nature of the effective force directly depends on the values of $aE$ and $L$, i.e., the first term will be repulsive, while the second term will be attractive if $aE < L$ and vice versa. Figure \ref{EF}, describes the behaviour of effective force at different values of spin and deviation parameter $g$, of the BH. The graphical representation shows stronger attraction of the effective force on photons in case of fast rotating BHs (upper left panel). Moreover, the effective force also gets increases, as the values of $g$ increases (upper right panel). The bottom panel of Fig. \ref{EF}, shows the comparison of effective force on photons around a rotating regular Hayward, Kerr and Schwarzschild BHs. It can be observed that in compared to the Kerr and Schwarzschild BHs, the effective force is more attractive for the regular rotating Hayward BH while having a minimum attraction for Schwarzschild BH.
\section{The Penrose mechanism}
\label{sec:4}
This section aims to inspect the Penrose mechanism within ergoregion of a regular rotating Hayward BH, which is among the attractive and important problems of GR. The primary goal of the mechanism is to extract energy from rotating BHs, it depends on the conservation of energy and momentum. It is one of the accurate and attractive energy extraction technique as compared to those of the nuclear reactions. The necessary and sufficient conditions for energy extraction from rotating BH via the said mechanism is the absorption of angular momentum and negative energy. Our main focus will be on studying the effects of deviation parameter $g$ on both of the negative energy states $E$ and efficiency of the mechanism. Furthermore, we will also investigate the effects of rotational parameter $a$, on efficiency as well as on negative energy $E$, of the BH.
\subsection{Negative energy states}
Negative energy states within ergosphere of a BH has essential consequences in BH physics. They could be occurred because of the counter-rotating orbits (as in Kerr BH) and may also because of the electromagnetic interaction (as in RN BH). 
The radial equation of motion \eqref{radial}, can be rewritten as
\begin{eqnarray}\label{negenergy}
	E^{2}[(r^{2}+a^{2})r^2 + 2a^{2}mr]- 2aEL(2mr)+ L^{2}(2mr-r^2)+r^2 \Delta\epsilon=0.
\end{eqnarray}
It is seen from Eq.(\ref{negenergy}) that there are two roots of the equation with respect to $E$ and $L$ which can be obtained as
\begin{eqnarray}
	\label{NE0}
	E_\pm&=&\frac{2aLmr \pm \sqrt{[L^{2}r^{4}-(r^{4}+a^{2}(2mr+r^2))r^2\epsilon] \Delta }} {r^{4}+a^{2}(2mr+r^2)},\label{NE1}
	\\ L_\pm&=&\frac{2aEmr\pm \sqrt{[E^{2}r^{4}-(2mr-r^2)r^2 \epsilon]\Delta}} {(2mr-r^2)}.
\end{eqnarray}
We have made use of the following identity to obtain the above results
\begin{eqnarray}
	\label{NE2}
	\Delta r^{4} - a^{2}(2mr)^{2}=(r^{4}+a^{2}(r^{2}+2mr))(r^{2}-2mr).
\end{eqnarray}

\begin{figure*} \vspace{-0.5cm}
	\includegraphics[width=.482\textwidth]{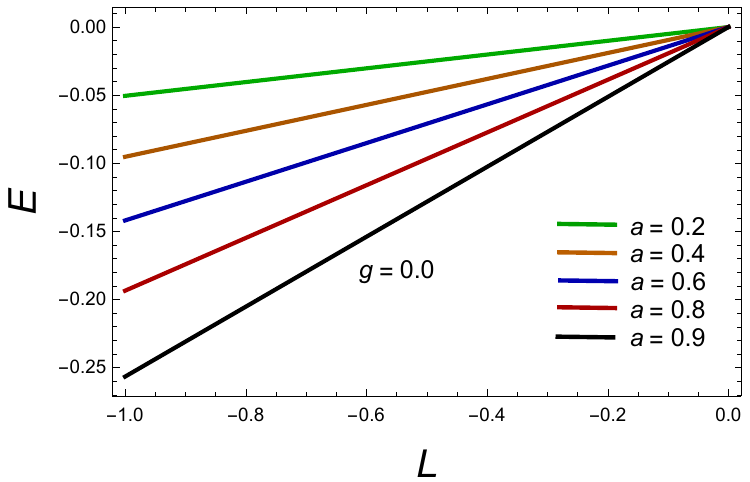}
	\includegraphics[width=0.482\textwidth]{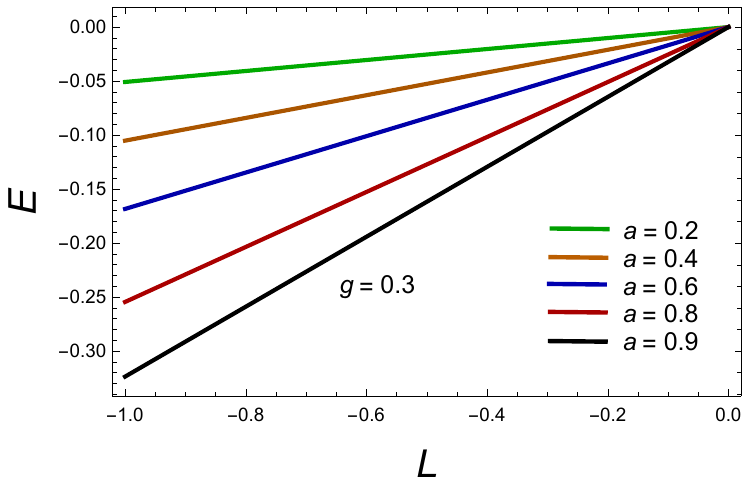}
	\centering
	\includegraphics[width=0.482\textwidth]{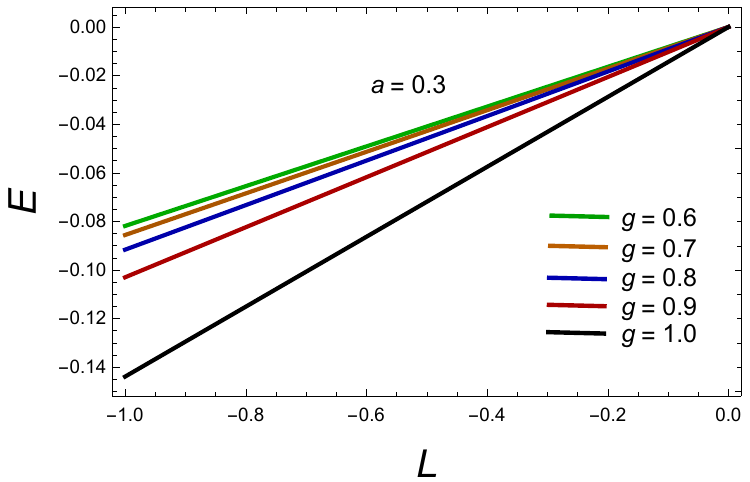}
	\caption{Graphical behaviour of negative energy states $E$, versus the angular momentum $L$.}\label{NE01}
\end{figure*}
Equation \eqref{NE1}, can be utilized to deduce the condition of negative energy states. Initially, we assume $E=1$ and consider only $+$ sign of Eq. (\ref{NE1}) satisfies the assumption. For $E<0$, it is also necessary that $L<0$, and
\begin{eqnarray}\label{NE3}
	\left(2a L m r\right)^{2}>[L^{2}r^{4}-(r^{4}+a^{2}\left(r^{2}+2mr\right) r^2 \epsilon]\Delta.
\end{eqnarray}

With the help of Eq. (\ref{NE2}), the above inequality (\ref{NE3}) can be rewritten as
\begin{eqnarray}\label{NE4}
	\left[(r^{2}-2mr)L^{2}-\epsilon \Delta r^{2}\right]\left[a^{2}\left(r^{2}+2mr\right)+r^{4}\right]<0.\end{eqnarray}

From the above inequality, it could be found that $E<0 $ if and only if $L<0$,
and
\begin{eqnarray}
	\left(\frac{r^2-2mr}{r^2}\right)<\frac{\Delta}{L^{2}}{\epsilon}.
\end{eqnarray}
On substituting the deviation parameter $g=0$, the aforementioned inequality reduces to the case of Kerr BH \cite{Chandrasekhar}. A graphical representation of the angular momentum $L=L_+$ , versus negative energy $E=E_+$, is shown in Fig.  \ref{NE01}. The top panel of the graph indicates that in comparison with the standard Kerr BH, the regular rotating Hayward BH have more negative energy. In addition, the BH spin results in an increase of its negative energy. It could also be noted from the lower panel of the plot that, for a fixed value of the spin parameter $a$,  the deviation parameter $g$ contribute to the negative energy, which in fact behave like the tidal charge parameter of the braneworld Kerr BH \cite{saeed}. This shows the strong influence of deviation parameter $g$ on negative energy $E$, of the BH.

The mechanism of energy extraction from rotating BHs leads to the irreducible mass of the BH. At the point when a particle with negative energy goes into the BH, the mass of the BH changes by an 
amount $\delta_M =E$ \cite{Abdujabbarov1}. There is no upper bound on, $\delta_M$ as it can be increased by increasing mass of the injected particle. Nevertheless, there is a lower bound on $\delta_M$ and every incident particle posses negative energy decreases the mass of the BH until its irreducible mass. By making use of Eq. \eqref{NE0}, we can find the lower bound of irreducible mass $\delta_M$. The discriminant of Eq. \eqref{NE0} vanished at horizons and as a result, we acquired the lower bound of irreducible mass
\begin{eqnarray}
	\label{IM}
	\delta_M=\frac{2amL}{r_+^3+a^2(2m+r_+)}.
\end{eqnarray}

From the above equation, it is deduced that to extract energy from a BH, the incident particle must have negative angular momentum. Moreover, it is observed that both of the spin, and deviation parameter $g$ of the BH, has great influence on irreducible mass of the BH.
\subsection{Efficiency of the mechanism}
The efficiency of energy extraction with the help of Penrose mechanism has an essential consequence in the energetics of rotating BHs. The working steps of this technique include a particle having energy $E_{(0)}$ goes into ergoregion of a BH and subdivided into two sub-particles termed as 1 and 2, and, respectively, possess energy $E_{(1)}$ and $E_{(2)}$. The sub-particle 1 has greater energy in comparison with the injected one and leaves the ergosphere, while the sub-particle 2 having negative energy enters to the BH. Henceforth, utilizing the law of conservation of energy, we can write
\begin{equation}
	\nonumber
	E_{(0)}=E_{(1)}+E_{(2)},
\end{equation} 

The consequences of sub-particle 2, i.e., $E_{(2)}<0$, results in $E_{(1)} > E_{(0)}$. Let us assume that, $\nu=dr/dt$ be the radial velocity of a particle with respect to an observer at infinity. Henceforth, by the laws of conservation of energy, as well as angular momentum, we have
\begin{eqnarray}
	\label{Eff1}
	L=p^{t}\Omega \,, \quad E=-p^{t}\mathcal{Z},
\end{eqnarray}

where
\begin{equation}
	\mathcal{Z} \equiv g_{tt} + g_{t\phi}\Omega.
\end{equation}

Since $p^{\eta} p_{\eta}=-m^{2}$, then
\begin{equation}
	\label{Eff2}
	g_{tt}\dot{t}^{2}+2g_{t\phi}\dot{t}\dot{\phi}+g_{rr}\dot{r}^{2}+g_{\phi\phi}\dot{\phi}^{2}=-m^{2}.
\end{equation}

Multiplying Eq. (\ref{Eff2}) with $1/\dot{t}^{2}$, we get
\begin{equation}
	\label{Eff3}
	g_{tt}+2 \Omega\ g_{t\phi}+\Omega^{2}g_{\phi\phi} +\frac
	{\upsilon^{2}}{\Delta}r^{2}=-\left(\frac{m\mathcal{
			Z}}{E}\right)^{2}.
\end{equation}

As in the left-hand side, the fourth term of Eq. \eqref{Eff3} is always positive, whereas, its right-hand side is equal to zero or negative. This makes us able to write Eq. \eqref{Eff3}, in the following form
\begin{eqnarray}
	\label{Eff4}
	g_{tt}+2 \Omega\ g_{t\phi}+\Omega^{2}g_{\phi\phi}=-\frac {\upsilon^{2}}{\Delta}r^{2}-\left(\frac{m
		\mathcal{Z}}{E}\right)^{2}\leq 0.
\end{eqnarray}

The angular velocity should assure the constraints
of $\Omega_{-}\leq\Omega\leq\Omega_{+}$ \cite{Lightman}, where
\begin{eqnarray}\nonumber
	\Omega_{\pm}=\frac{ -g_{t\phi}\pm\sqrt{g_{t\phi}^2-g_{tt}g_{\phi\phi}}}{g_{\phi \phi}}.
\end{eqnarray}
Utilizing Eq. \eqref{Eff1}, the association of conservation of energy and angular momentum can be communicated as
\begin{equation}
	\label{Eff5}
	p^{t}_{(0)}\mathcal{Z}_{(0)}=p^{t}_{(1)}\mathcal{Z}_{(1)}+p^{t}_{(2)}\mathcal{Z}_{(2)},
\end{equation}

\begin{equation}
	\label{Eff6}
	p^{t}_{(0)}\Omega_{(0)}=p^{t}_{(1)}\Omega_{(1)}+p^{t}_{(2)}\Omega_{(2)}.
\end{equation}

Therefore, the efficiency $(\eta)$ of Penrose mechanism could be described as
\begin{equation}
	\eta=\frac{E_{(1)}-E_{(0)}}{E_{(0)}}=\mathcal{Y}-1, 
\end{equation}

here $\mathcal{Y}>1$ as $\mathcal{Y}=E_{(1)}/E_{(0)}$. From Eqs. (\ref{Eff1}), (\ref{Eff5}) and \eqref{Eff6}, we obtain

\begin{equation}
	\label{Eff7}
	\mathcal{Y}=\frac{E_{(1)}}{E_{(0)}}=\frac{(\Omega_{(0)}-\Omega_{(2)})\mathcal{Z}_{(1)}}
	{(\Omega_{(1)}-\Omega_{(2)})\mathcal{Z}_{(0)}}.
\end{equation}

Here, we assume an injected particle goes into the ergosphere of the BH with energy $E_{(0)}=1$, breaks into two photons bearing momenta $p_{(1)}=p_{(2)}=0$. From Eq. \eqref{Eff7} we note that efficiency of the process can be maximized, if $\Omega_{(1)}$ has its smallest and $\Omega_{(2)}$ has its greatest values simultaneously, which needs $\upsilon_{(1)}=\upsilon_{(2)}=0$. Henceforth, it implies
\begin{eqnarray}\label{Eff8}
	\Omega_{(1)}=\Omega_{+}, \quad \Omega_{(2)}=\Omega_{-}.
\end{eqnarray}
As a result, the parameter $\mathcal{Z}$ simplifies to
\begin{eqnarray}\label{Eff9}
	\mathcal{Z}_{(0)}=g_{tt}+\Omega_{(0)} \ g_{t\phi}, \quad
	\mathcal{Z}_{(2)}=g_{tt}+\Omega_{-} \ g_{t\phi}.
\end{eqnarray}
The four-momenta of pieces are
\begin{eqnarray}\nonumber
	p_{\eta}=p^{t}(1,0,0,\Omega_{\kappa}),\quad \kappa=1,2.
\end{eqnarray}
Therefore, Eq. \eqref{Eff2} will take the form
\begin{eqnarray}\label{Ef}
	(g_{t\phi}^{2}+g_{\phi\phi})\Omega^{2} + 2\Omega(1+g_{tt})g_{t\phi}+(1+g_{tt})g_{tt}=0.
\end{eqnarray}
Henceforth, by solving the aforementioned equation, we can obtain the corresponding angular velocity of the injected particle as
\begin{eqnarray}
	\Omega_{(0)}=\frac{-(1+g_{tt})g_{t\phi}+\sqrt{(1+g_{tt})(g^{2}_{t\phi}-g_{\phi\phi}g_{tt})}}{g^{2}_{t\phi}+g_{\phi\phi}}.
\end{eqnarray}
Subsequently, using the values of Eqs. \eqref{Eff8} and \eqref{Eff9} in \eqref{Eff7}, efficiency of the energy extraction modifies to
\begin{eqnarray}
	\eta=\frac{(g_{tt}+g_{t\phi} \Omega_{+})(\Omega_{(0)}-\Omega_{-})}
	{(g_{tt}+g_{t\phi}\Omega_{0})(\Omega_{(+)}-\Omega_{-})}-1.
\end{eqnarray}
In order to obtain maximum efficiency ($\eta_{max}$) of the extracted energy, it is mandatory for the incoming particle to be split at the horizon of BH. Subsequently, the above equation simplifies to
\begin{eqnarray}
	\eta_{max}= \frac{1}{2}\left(\sqrt{\frac{2m}{r_{+}}}-1\right).
\end{eqnarray}
\begin{table*}
	\begin{center}
		\textbf{Table 3:} Maximum efficiency $\eta_{max}$ $(\%)$ of
		the energy extraction via the Penrose mechanism.
			\begin{tabular}{c c c c c c c c c }
				\hline \hline \noalign{\smallskip\smallskip}
				&  {a=0.3}  &  {a=0.5}  &  {a=0.7}  &   {a=0.8}  &  {a=0.9}  &  {a=0.99}  &  {a=1.0}& \\ \noalign{\smallskip}
				\hline \noalign{\smallskip\smallskip}
				$g=0.0$     & 0.5859 & 1.7638 & 4.0084 & 5.9017 & 9.0098  & 16.1956 & 20.7107&\\
				$g=0.2$     & 0.5872 & 1.7688 & 4.0257 & 5.9386 &  9.1161 & 17.8565 & &\\
				$g=0.4$     & 0.5967 & 1.8050 & 4.1563 & 6.2295 & 10.1143 & & & \\
				$g=0.6$     & 0.6257 & 1.9199 & 4.6326 & 7.6708 & & & & \\
				$g=0.7$     & 0.6544 & 2.0424 & 5.3504 & & &  & & \\
				$g=0.8$     & 0.7019 & 2.2743 &  & & &  & & \\
				\hline\hline \noalign{\smallskip}
			\end{tabular}
	\end{center}\label{ep}
\end{table*}
\begin{figure*} \vspace{-0.0cm}
	\includegraphics[width=.482\textwidth]{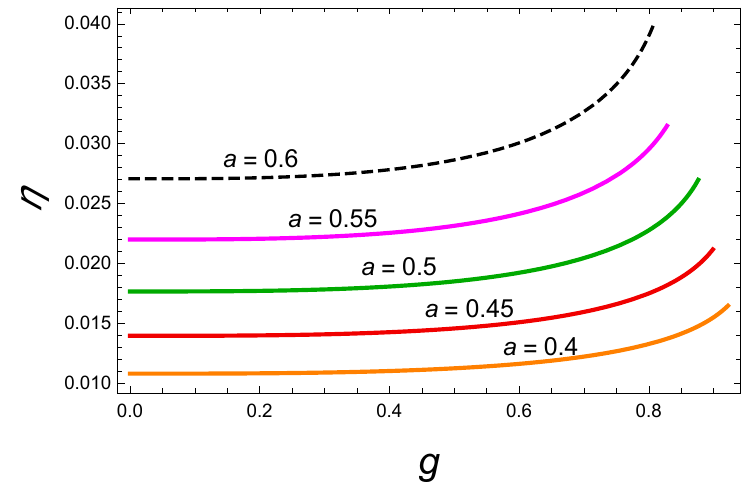}
	\includegraphics[width=0.482\textwidth]{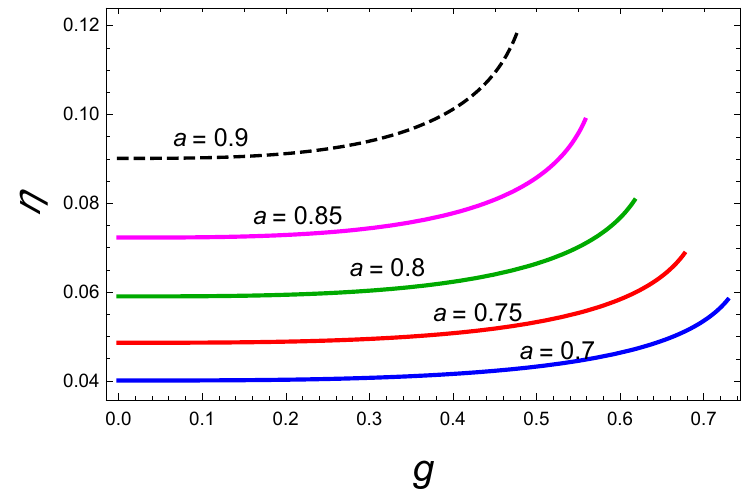}
	\centering
	\includegraphics[width=0.482\textwidth]{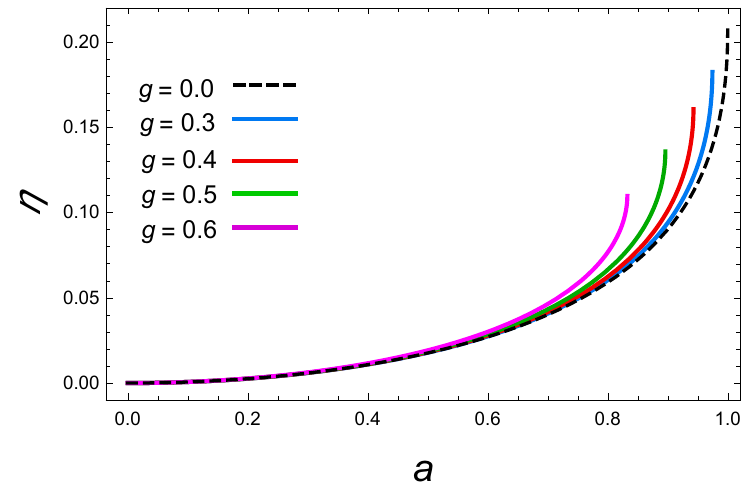}
	\caption{Maximum efficiency of the energy extraction versus $g$ (upper panel), whereas $a$ (lower panel).}\label{EE1}
\end{figure*}
Table {\bf{3}} shows the numerical calculation of maximum efficiency of the energy extraction mechanism at various values of the spin and deviation parameter $g$, of the BH. It is observed that efficiency of the mechanism increases as the values of $g$ increases. For $g=0$ and $a=1$, efficiency can be maximum, i.e., $20.7\%$, which is the limiting value for the extreme Kerr BH \cite{Chandrasekhar}, denoted by black dotted curves in the lower panel of the plot in Fig \ref{EE1}. Graphical behaviour of maximum efficiency of the energy extraction with the help of Penrose mechanism is portrayed in Fig. \ref{EE1}, at various choices of the deviation and rotation parameter $a$. We observed that for increasing values of $a$, as well as $g$, efficiency of the process increases. In current work, the efficiency behaviour of energy extraction is identical to that of the braneworld Kerr BH, but it can be noted that in comparison with deviation parameter $g$, the efficiency increase rate is higher for the brane parameter of the BH \cite{saeed}.
\section{Concluding remarks}
\label{sec:5}
In this manuscript, we have explored the Penrose mechanism of energy extraction together with neutral particle motion around a regular rotating Hayward BH. Geodesics and more especially the circular geodesics are essential for the studying particles motion and the dynamics of galaxies. We figured out the geodesic equation of particles orbiting a regular rotating Hayward BH. We have graphically analysed the effect of spin, as well as deviation parameter $g$, on neutral particle dynamics. Moreover, we have explored the ergoregion, static limit and event horizon of the BH. The obtained results showed that both spin and deviation parameters, of the BH, contributes to the thickness of ergoregion.

While studying the circular geodesics, we have obtained both of the innermost stable circular and photon orbits. We found that $g$ result in diminishing the photon orbits of both prograde, as well as the retrograde motion of particles. On the other hand, BH rotation decreases the photon orbits of prograde motion, while contributes to the photon orbits of retrograde motion. Moreover, we have made use of the effective potential to investigate the stability of circular orbits. We observed an increasing behaviour of the effective potential for higher values of $a$, which summarizes that fast-rotating BHs leads to a more stable circular orbit. Besides, the stability of circular orbits around a regular rotating Hayward BH is compared to that of Kerr and Schwarzschild BHs and concluded that the effective potential is more stable, in case of regular rotating Hayward BH in comparison with the Kerr and Schwarzschild BHs.

We have utilized the Penrose mechanism to explore the process of energy extraction from a regular rotating Hayward BH. While on investigating the negative energy states we have figured out that, only in case of non-positive angular momentum, the negative energy states are possible.  Moreover, we concluded that both $a$ and $g$ contributes to the negative energy states of the BH. We observe that the standard Kerr BH possess less amount of the negative energy as compared to the regular rotating Hayward BH. Furthermore, from the study of energy extraction, it is observed that the efficiency of energy extraction increases with both spin and deviation parameters of the BH. The BH rotation has an extraordinary impact on the energy extraction, as in the case of fast-rotating BHs, particles can take energy from BH rotation consequently, the energy extraction from a rotating BH increases. Our finding reflects considerable resemblance to that of Kerr BH, as in both cases the rotational parameter notably affects the efficiency of energy extraction. The obtained results show that, as compared to the Kerr BH, one can extract more energy from a regular rotating Hayward BH.
\subsubsection*{Acknowledgment:}
Jingli Ren would like to acknowledge the financial support provided for this research via the Open Fund of State Key Laboratory of Power Grid Environmental Protection (No. GYW51202101374), the National Natural Science Foundation of China (52071298), Zhong Yuan Science and Technology Innovation Leadership Program (214200510010). They also thank the reviewers for their valuable reviews and kind suggestions.

\end{document}